\documentclass[onecolumn, showpacs,superscriptaddress,nofootinbib,floatfix, amsfonts, aps]{revtex4}

\usepackage{amsmath}
\usepackage{bm}
\usepackage{graphicx}
\usepackage{mathrsfs}
\usepackage{footmisc}
\usepackage{multirow}
\usepackage{graphicx}
\usepackage{booktabs, ctable}
\usepackage{xcolor, color, framed}

\begin{document}

\title{Charmonium Coherent Photoproduction and Hadroproduction with 
Effects of Quark Gluon Plasma}

\author{Wei Shi}
\affiliation{Department of Physics, Tianjin University, Tianjin 300350, China}
\author{Wangmei Zha}
\affiliation{Department of Modern Physics, University of Science and Technology of China, Hefei 230026, China }
\author{Baoyi Chen}
\thanks{Corresponding author: baoyi.chen@tju.edu.cn }
\affiliation{Department of Physics, Tianjin University, Tianjin 300350, China}
\affiliation{Institut f\"ur Theoretische Physik, Goethe-Universit\"at Frankfurt,
Max-von-Laue-Str. 1, D-60438 Frankfurt am Main, Germany}
\date{\today}

\begin{abstract}
We study the charmonium coherent photoproduction and hadroproduction 
consistently with modifications from both cold and hot nuclear matters. 
The strong electromagnetic fields from fast moving nucleus interact with 
the other target nucleus, producing abundant charmonium in the 
extremely low transverse 
momentum region $p_T<0.1$ GeV/c. This results in significative enhancement of 
$J/\psi$ nuclear modification factor in semi-central and peripheral collisions. 
In the middle $p_T$ region such as $p_T<3\sim 5$ GeV/c, $J/\psi$ final yield 
is dominated by the combination process of single charm and anti-charm quarks 
moving in the deconfined matter, 
$c+\bar c\rightarrow J/\psi +g$. In the higher $p_T$ region, 
$J/\psi$ production are mainly from parton initial hard scatterings at the 
beginning of 
nucleus-nucleus collisions and decay of B hadrons. We include all of these 
production mechanisms and explain the experimental data well 
in different colliding centralities and transverse momentum regions.      
\end{abstract}
\maketitle

With the nuclear collisions at the relativistic heavy ion collisions (RHIC) and 
the large hadron collider (LHC), there have been a lot of interesting topics 
about nuclear properties studied in experiments and theories. One of the main 
goals at RHIC and LHC is to find a deconfined state of nuclear matter, called 
``quark-gluon plasma'' (QGP), which may be produced in the extremely high 
energy and/or baryon densities with a phase transition~\cite{Bazavov:2011nk}, 
and furthermore, extract the transport 
properties of this QGP~\cite{Song:2008hj}. There are also other projects 
referred to as ``\emph{non-QGP}'' 
physics, concerning about cosmic ray physics and others~\cite{Anselm:1989pk}. As 
QGP can only be produced in the nucleon collisions in the overlap area of two 
nuclei, ``\emph{non-QGP}'' topics are usually studied in Ultra-peripheral nuclear 
collisions (UPC) where QGP background is absent. 

In order to study the existence of the extremely hot deconfined matter, 
produced in the early stage of heavy ion collisions, 
$J/\psi$ abnormal suppression has been proposed as one of 
sensitive signals by Matsui and Satz in 1986~\cite{matsui}. 
$J/\psi$ suffers relatively weaker dissociation by the 
hadron gas compared with QGP, due to its large binding energy~\cite{Satz:2005hx}. 
With strong color 
screening effect and parton inelastic scatterings of QGP, $J/\psi$ production 
can be significantly suppressed in nucleus-nucleus collisions, which has been 
observed in many experiments at RHIC and LHC colliding energies in semi-central 
and central collisions (with impact parameter $b<2R_A$, where $R_A$ is the 
nuclear radius)~\cite{Adare:2006ns,Abelev:2009qaa,Abelev:2013ila,Adam:2016rdg}. 
From RHIC to LHC, $J/\psi$ production is relatively enhanced in 
the transverse momentum region $p_T\le 3\sim5$ GeV/c compared with the scale of the 
pp collisions~\cite{Abelev:2013ila}. This phenomenon 
has been well explained with charmonium 
regeneration mechanism: At higher colliding energy, 
more charm quark pairs can be produced from hard process 
in hadronic collisions, which enhance the recombination probability 
of charm and anti-charm quarks inside 
QGP~\cite{Thews:2000rj,BraunMunzinger:2000px,
Yan:2006ve,Andronic:2003zv,Zhao:2011cv}. 
New $J/\psi$s are continuously 
regenerated during the QGP evolutions. 
With sufficiently high initial temperature 
of QGP and abundant number of charm quarks, the primordially produced $J/\psi$s 
are strongly suppressed and therefore regenerated $J/\psi$s dominate the final 
total yields in Pb-Pb collisions at LHC energies~\cite{Zhou:2014kka,Chen:2015iga}.

Also, in the relativistic heavy ion collisions, the nuclei with charges 
move with nearly the 
speed of light, $v>0.99c$ at RHIC and LHC. The strongest magnetic field 
on earth can be produced, with 
a magnitude of $eB\sim 5m_\pi^2$ at RHIC Au-Au and 
$70m_\pi^2$ at LHC Pb-Pb collisions~\cite{Deng:2012pc,Tuchin:2013ie}. 
Electromagnetic fields become strongly Lorentz contracted in the 
longitudinal direction (nuclear accelerating direction, defined as $z-$ axis) 
~\cite{Krauss:1997vr}. 
In semi-central nuclear collisions, both strong electromagnetic fields 
and the QGP can be produced~\cite{Deng:2012pc, Shen:2012vn}. 
In the QGP, there will be interactions between magnetic fields and chiral light 
quarks at the limit of zero mass $m_q=0$ in the deconfined matter. 
A lot of topics about the magnetic field induced QCD chirality are studied 
widely, such as chiral magnetic effect~\cite{Metlitski:2005pr,Kharzeev:2007jp}, 
chiral magnetic wave~\cite{Burnier:2011bf}, 
chiral vortical effect~\cite{Kharzeev:2010gr}, and 
chiral electric separation effect~\cite{Huang:2013iia}. 

The electromagnetic fields can also interact with the 
other nucleus ($\gamma A$ or $\gamma N$ interactions) 
or with the electromagnetic 
fields of the other nucleus ($\gamma\gamma$ interactions), and produce hadronic 
final states~\cite{Krauss:1997vr,Baur:2001jj,Klein:2003vd,Bertulani:2005ru, 
Yu:2014eoa,Yu:2015kva}. 
Fermi first proposed 
that the transverse electromagnetic fields 
can be approximated as a swarm of equivalent photons, called 
``Equivalent Photon Approximation'' (EPA)~\cite{Fermi:1924tc}. This idea was also 
extended by Weizs\"acker~\cite{vonWeizsacker:1934nji} and 
Williams~\cite{Williams:ref} 
independently and therefore also 
called ``Weizs\"acker-Williams-Method''. 
This allows a simple and straightforward 
calculations of vector meson photoproduction
between the target nucleus 
and electromagnetic fields~\cite{Khoze:2002dc}. 
With long range of electromagnetic interactions, 
a hard equivalent photon from one nucleus may penetrate into the other nucleus 
and interact with quarks or gluons. Therefore, 
one goal of this photoproduction is to study the parton densities 
of a bound nucleon inside the nucleus, such as shadowing effect.  
The interactions of $\gamma A$ (or $\gamma N$) and $\gamma\gamma$ can produce
heavy quark pairs $Q\bar Q$ , dileptons $l\bar l$, and vector mesons 
$V=\phi, \rho^0, J/\psi, \psi(2S)$~\cite{Klein:1999qj,Ducati:2013tva,
Ducati:2013bya, Yu:2013uka}. 
In semi-central collisions with $b<2R_A$,  
these heavy quarks or vector mesons will also 
go through hot medium and suffer 
dissociations~\cite{Klusek-Gawenda:2015hja}. 
With the strong electromagnetic 
fields, charmonium photoproduction may become larger than the hadroproduction in 
extremely low $p_T$ region 
even in the semi-central collisions with the production of deconfined matter, 
which has already been
observed by experiments at RHIC~\cite{Zha:2017etq} and LHC~\cite{Adam:2015gba}. 

In Pb-Pb semi-central collisions 
at $\sqrt{s_{NN}}=2.76$, 
the initial temperature of QGP is around $(1.5\sim2)T_c$ where $T_c$ is the critical 
temperature of deconfined phase transition~\cite{Chen:2015iga}. 
With the realistic evolutions of QGP and all the 
sources of charmonium production 
including photoproduction and hadroproduction (consists of primordial production, 
regeneration and decay of B hadrons), 
we give the $J/\psi$ nuclear modification 
factor as a function of number of participants $R_{AA}(N_p)$ and 
transverse momentum $R_{AA}(p_T)$. We find that primordial production, regeneration 
and photoproduction plays the important role in different $p_T$ regions of 
charmonium production, 
showing different physics on heavy ion collisions. 

With a large mass, charmonium evolutions in the hot medium can be described by a 
classical Boltzmann transport equation. It has described well hadroproduced 
charmonium $R_{AA}(N_p)$
at different colliding energies from 
SPS to LHC~\cite{Zhu:2004nw,Chen:2015ona,Zhou:2014kka}, 
mean transverse momentum 
square $\langle p_T^2\rangle(N_p)$~\cite{Liu:2009wza}, 
and rapidity distribution $R_{AA}(y)$~\cite{Chen:2015iga}. The transport 
equation for hadroproduced charmonium with cold and hot nuclear matter effects 
is    
\begin{align}
{\partial f_\Psi \over \partial \tau} + {\bf v}_\Psi\cdot {\bf \nabla} 
f_\Psi = -\alpha_\Psi f_\Psi +\beta_\Psi
\label{tran}
\end{align} 
where $f_{\Psi}$ is the charmonium 
distribution in phase space. 
$\tau=\sqrt{t^2-z^2}$ is the local proper time 
(here $t$ is the time variable). The second 
term on the L.H.S. of Eq.(\ref{tran}) represents free streaming of $\Psi$ 
with transverse velocity 
${\bf v}_T={\bf p}_T/\sqrt{m_\Psi^2 +p_T^2}$. 
On the R.H.S. of Eq.(\ref{tran}), the loss term $\alpha_\Psi$ 
represents charmonium decay rates in QGP due to color screening effect 
and parton inelastic scatterings, and is written as  
\begin{equation}
\label{Cdfactor}
\alpha_\Psi ={1\over 2E_T} \int {d^3{\bf k}\over {(2\pi)^32E_g}}\sigma_{g\Psi}({\bf p},{\bf k},T)4F_{g\Psi}({\bf p},{\bf k})f_g({\bf k},T)
\end{equation}
where $E_g$ and $E_T=\sqrt{m_\Psi^2 +p_T^2}$ are the gluon energy and charmonium 
transverse energy, respectively. 
$F_{g\Psi}$ is the flux factor. 
Charmonium decay rate in QGP is proportional to 
the gluon thermal density $f_g({\bf k},T)$ 
and also their inelastic cross sections $\sigma_{g\Psi}(T)$~\cite{Liu:2009wza}. 
The cross section for gluon dissociation in vacuum $\sigma_{g\Psi}(0)$ can be 
derived through the operator production expansion. It is extended to the 
finite temperature by geometry scale, $\sigma_{g\Psi}(T)=\sigma_{g\Psi}(0)\times 
\langle r_\Psi^2\rangle(T)/\langle r_\Psi^2\rangle(0)$, where $\langle r_\Psi^2
\rangle(T)$ is the charmonium mean radius square at finite temperature, which 
can be obtained from potential model with the 
color screened heavy quark potential from Lattice calculations~\cite{Chen:2017duy}. 
The divergence of 
charmonium radius at $T\rightarrow T_d^\Psi$ indicates the melting of the bound 
state $\Psi$. Charm and anti-charm quarks in the deconfined matter can also 
combine to generate a new bound state, represented by the 
gain term $\beta_\Psi$. It is connected 
with the loss term $\alpha_\Psi$
through detailed balance between the gluon dissociation process and its 
inverse process, $g+\Psi\leftrightarrow c+\bar c$. 

With the loss and gain terms in Eq.(\ref{tran}), 
one can write the analytic solution for charmonium phase space distribution at 
the time $\tau$ to be
{\footnotesize{
\begin{align}
f_\Psi({\bf p}_T,{\bf x}_T, \tau|{\bf b}) 
=
&f_\Psi({\bf p}_T,{\bf x}_T-{\bf v}_\Psi(\tau-\tau_0),\tau_0) 
e^{-\int_{\tau_0}^{\tau} d\tau^{\prime}\alpha_\Psi({\bf p}_T,
{\bf x}_T -{\bf v}_\Psi(\tau-\tau^\prime),\tau^\prime)} \nonumber \\
&+\int_{\tau_0}^{\tau}d\tau^\prime \beta_\Psi({\bf p}_T,
{\bf x}_T -{\bf v}_\Psi(\tau-\tau^\prime),\tau^\prime)
\times 
e^{-\int_{\tau^\prime}^{\tau} d\tau^{\prime\prime}\alpha_\Psi({\bf p}_T,
{\bf x}_T -{\bf v}_\Psi(\tau-\tau^{\prime\prime}),\tau^{\prime\prime})}
\label{eq-analy}
\end{align} 
}}%%
where $\tau_0\sim 0.6 fm/c$ is the time scale 
of QGP reaching local equilibrium, fixed by light hadron spectra in hydrodynamic 
models. 
Charmonium initial distribution in nucleus-nucleus collisions 
$f_\Psi({\bf p}_T, {\bf x}_T,\tau_0|{\bf b})$ 
is obtained by the geometry scale with pp collisions 
${\bar f}_\Psi({\bf p}_T, {\bf x}_T,\tau_0|{\bf b})$ 
in Eq.(\ref{eq-initf}), with additional 
modifications from shadowing effect~\cite{Eskola:1998df} and 
Cronin effect~\cite{Chen:2015iga}.  
{\footnotesize 
\begin{align}
\label{eq-initf}
{\bar f}_\Psi({\bf p}_T,{\bf x}_T,\tau_0|{\bf b}) 
= \int dz_A dz_B 
\rho_A ({\bf x}_T +{{\bf b}\over 2},z_A) \rho_B ({\bf x}_T -{{\bf b}\over 2},z_B) 
{d^2\sigma_{J/\psi}^{pp}\over dy 2\pi p_T dp_T}
\end{align}
}%%
where $\rho_{A(B)}$ is the Woods-Saxon nuclear density. 
The differential cross section for charmonium hadroproduction in pp collisions 
is parametrized with~\cite{Chen:2015iga,Chen:2016dke},
{\small
\begin{align}  
\label{eq-ppJpsi}
{d^2\sigma^{pp}_{J/\psi}\over dy 2\pi p_T dp_T} = {2(n-1)\over 2\pi (n-2)
\langle p_T^2\rangle }[1+{p_T^2\over (n-2)\langle p_T^2\rangle }]^{-n} 
{d\sigma_{J/\psi}^{pp}\over dy}
\end{align}
}%%
Here $y=1/2\ln[(E+p_z)/(E-p_z)]$ and $\langle p_T^2\rangle$ are the rapidity 
and the mean transverse momentum square of $J/\psi$. 
At $\sqrt{s_{NN}}=2.76$ TeV pp collisions, we fit the 
experimental data of $J/\psi$ inclusive 
hadroproduction cross section at forward rapidity $2.5<y<4$ to obtain 
$n=4.0$ and 
$\langle p_T^2\rangle =7.8\ \rm{(GeV/c)^2}$~\cite{Zhao:2017yan}. 
$J/\psi$ rapidity differential cross section is 
$d\sigma_{J/\psi}^{pp}/dy=2.3\ \mu b$ in the forward rapidity~\cite{Aaij:2012asz}.

The regeneration rate $\beta_\Psi$ is 
proportional to the densities of charm and anti-charm 
quarks which are produced through nuclear hard process. 
Charm quark initial densities in 
nucleus-nucleus collisions are obtained through
{\small 
\begin{equation}
\label{dnsty}
\rho_c({\bf x}_T, \eta,\tau_0) = {{d\sigma^{pp}_{c{\bar c}}} 
\over d\eta}{ T_A({\bf x}_T+{\bf b}/2)T_B({\bf x}_T-{\bf b}/2)\cosh(\eta)\over \tau_0}
\end{equation} }%%
where $T_{A(B)}({\bf x}_T)$ is the thickness function of nucleus A(B) at the 
transverse coordinate ${\bf x}_T$, with the definition $T({\bf x}_T)=\int dz 
\rho({\bf x}_T,z)$. $\eta = 1/2\ln[(t+z)/(t-z)]$ is the spatial rapidity. 
Charm pair production cross section is taken to be  
$d\sigma_{c\bar c}^{pp}/d\eta=0.38$ mb in 
the forward rapidity $2.5<y<4$ of pp collisions at 2.76 TeV 
~\cite{Abelev:2012vra,Zhou:2014kka}. 
Shadowing effect reduces charm number by 20\% 
in Pb-Pb collisions~\cite{Eskola:1998df}.  
Inspired by the 
fact that D meson elliptic flows are comparable with light hadrons 
at $\sqrt{s_{NN}}=2.76 $ TeV Pb-Pb collisions~\cite{ALICE:2012ab,Abelev:2013lca}, 
we assume that charm quarks reach kinetic equilibrium at $\tau_0$. 
With the thermal production of charm pairs strongly suppressed by 
its large mass, evolutions of charm quark density $\rho_c({\bf x}_T,\eta, \tau)$ 
in the expanding QGP 
satisfy the conservation equation, 
\begin{align}
\partial_\mu(\rho_c u^\mu)=0
\end{align} 
where $u^\mu$ is the four velocity of QGP fluid cells, 
given by Eq.(\ref{eq-hydro}).   

The experimental data of charmonium $R_{AA}$ 
also includes the contribution from B hadron decay 
(called ``non-prompt'' $J/\psi$), with 
a fraction of $f_B=N_{B\rightarrow J/\psi}/N_{\rm{incl}}
=0.04+0.023p_T/(\rm{GeV/c})$ 
in the final inclusive yields, depending on the $p_T$ bins. 
B hadrons exist as bottom quarks in QGP and suffer energy 
loss~\cite{He:2014epa,He:2014cla,Yao:2017fuc}. 
This will change momentum distributions of bottom quark and B hadrons 
with the total 
number conservation~\cite{Chen:2013wmr}. 
One should treat the prompt and non-prompt $J/\psi$ 
separately. The prompt $J/\psi$ cross section is the product of inclusive cross 
section Eq.(\ref{eq-ppJpsi}) and the prompt fraction $(1-f_B)$, and taken as 
an input in Eq.(\ref{eq-initf}). 
In the extremely low $p_T$ region of $p_T<0.1$ GeV/c, the fraction of non-prompt 
$J/\psi$ drops to only around $0.04$. In a certain $p_T$ window, 
this non-prompt yield can 
be enhanced/suppressed due to bottom quark energy 
loss in QGP, which changes $J/\psi$ 
inclusive $R_{AA}$ in Eq.(\ref{eqRAAincl}) and will be discussed in 
details below.

The quark gluon plasma produced at RHIC and LHC turns out to be a strong 
coupling system. Its evolution can be simulated with (2+1) dimensional 
ideal hydrodynamic equations, with the 
assumption of Bjorken expansion in the longitudinal direction, 
\begin{align}
\label{eq-hydro}
\partial_\mu T^{\mu\nu}=0
\end{align}
where $T^{\mu\nu}=(e+p)u^\mu u^\nu -g^{\mu\nu}p$ is the energy-momentum tensor, 
$e$ and $p$ are the energy density and pressure. 
We also need the equation of state to solve above equations. 
The deconfined matter is taken to be an ideal gas of massless gluons, 
$u$ and $d$ quarks, and strange quark with mass $m_s=150$ 
MeV~\cite{Sollfrank:1996hd}.  
Hadron gas is taken to be an ideal gas of all known hadrons
and resonances with mass up to 2 GeV~\cite{Patrignani:2016xqp}. Two phases
are connected with first-order phase transition. 
With the initial energy density 
of QGP fixed by the charged hadron multiplicity from experiments, we give 
the initial maximum temperature of QGP in Table.\ref{TabQGP}.  
Both local temperature $T({\bf x}_T, \tau)$ and 
four velocity $u^\mu({\bf x}_T, \tau)$ of QGP depend on the coordinate ${\bf x}_T$ 
and the time $\tau$. 

\begin{table} [h]
\normalsize
\centering
\caption{Information of QGP based on (2+1) dimensional ideal 
hydrodynamics. $b$ and $N_p$ are the impact parameter and number of participants. 
$T_0^{\rm QGP}$ and $\tau_f^{\rm QGP}$ are the 
initial maximum temperature and lifetime of 
QGP. $T_c$ is the critical temperature of deconfined phase transition.
}
\label{TabQGP}
\vspace{0.15cm}
\begin{tabular}{|p{1.3cm}<{\centering}|*{5}{p{2.0cm}<{\centering}|}}
%{|p{1cm}|p{1cm}c|p{1cm}c|p{1cm}c|p{1cm}c|} % {m{100pt}<{\centering} m{100pt}<{\centering} m{100pt}<{\centering}}
\hline \multicolumn{4}{|c|}{  
Hydro in LHC $\sqrt{s_{NN}}$=2.76 TeV Pb-Pb, $2.5<y<4$}  \\
\hline
       b(fm) & $N_p$   &  $ T^{\rm QGP}_{0}/T_c$ & $\rm{\tau_{f}^{\rm QGP}}$ (fm/c) \\
%	based on (2+1)D ideal hydrodynamics  \\
\hline
0 &406   & $2.6$ & 7.3 \\
\hline
9 &124   & $2.1$ & 4.2 \\
\hline
10.2& 83 & $1.95$ & 3.5 \\
\hline 
12& 35 & $1.5$ & 2.3 \\
\hline
\end{tabular}
\end{table}

Experimental data about $J/\psi$ production shows significant 
enhancement 
in the extremely low $p_T$ region at $N_p<100$~\cite{Adam:2015gba}, which is 
attributed to additional photoproduction by the 
interactions between  
strong electromagnetic fields from one nucleus and the other target nucleus.
In table \ref{TabQGP}, 
one can see that even at the region of $N_p<120$, QGP initial temperature 
can still be $(1.5\sim 2)T_c$. Therefore, it is necessary to 
consider seriously both charmonium hadroproduction 
and photoproduction and also modifications from hot medium effects 
at the same time, to reach a full consistent conclusion. 

As discussed before, 
we employ the Equivalent Photon Approximation and 
treat the transverse electromagnetic fields as a swarm of quasi-real photons 
in the longitudinal direction. 
The photons fluctuate into a quark-antiquark pair which then scatter with 
the target nucleus, mediated by two gluons in the lowest order, but without 
the net color exchange. 
It is not clear whether the electromagnetic fields are from the entire 
source nucleus or spectators, and interact with the
entire target nucleus or target spectators at the time of nuclear 
collisions with $b<2R_A$~\cite{Zha:2017jch,Ducati:2017sdr}. 
We assume that the interactions between electromagnetic 
fields and target nucleus happen before hadronic collisions, and therefore 
the photoproduction do not consider the effects of nucleus broken process. 
In UPCs, the rapidity differential cross section of 
$J/\psi$ exclusive photoproduction has been well 
studied~\cite{Klein:2003vd,Guzey:2016piu}, 
\begin{align}
{d\sigma_{AA\rightarrow AAJ/\psi}\over dy}(y) =
{dN_{\gamma}\over dy}(y)\sigma_{\gamma A\rightarrow AJ/\psi}(y) 
+{dN_{\gamma}\over dy}(-y)\sigma_{\gamma A\rightarrow AJ/\psi}(-y) 
\label{eq-photoJpsi}
\end{align}
where $y$ is the rapidity of photoproduced vector mesons (in this case, $J/\psi$). 
The presence of two terms in Eq.(\ref{eq-photoJpsi}) indicates that the colliding 
nucleus can serve either as a source of electromagnetic fields or as a target. 
The rapidity of photoproduced $J/\psi$ is connected with photon energy as 
$y=\ln[2w/M_{J/\psi}]$ in laboratory reference frame, $M_{J/\psi}$ is the $J/\psi$ 
mass. Write the photon spectrum as a function of photon energy $w$, 
$dN_\gamma/dy = w\cdot dN_\gamma/dw$ with the relation of $dw/dy=w$. 
Now we extend the situation to all centralities with impact parameter 
dependence. In a certain centrality bin with $b_{min}\sim 
b_{max}$, the photon spectrum becomes~\cite{Klein:1999qj}, 
\begin{align}
n_\gamma(w) = &\int_{b_{min}}^{b_{max}}2\pi b db n_\gamma(w,b)\\
n_\gamma(w,b) =&
{1\over \pi R_A^2}
\int_0^{R_A} rdr\int_0^{2\pi}d\phi {d^3N_\gamma(w, b+r\cos(\phi))\over dwrdrd\phi}
\label{eq-phoden}
\end{align}
Since photons interact with the target nucleus coherently to produce vector mesons, 
the electromagnetic field strength is averaged over the surface of target nucleus 
with the area $\pi R_A^2$. 

With the photon-nucleus cross section and photon density, 
we write $J/\psi$ coherent photoproduction as a function of rapidity in 
AA collisions with 
impact parameter $b$ to be, 
\begin{align}
{dN_{J/\psi}\over dy}(y|b)=&w n_\gamma(w,b)\sigma_{\gamma A\rightarrow J/\psi A}(w) 
+ (y\rightarrow -y) 
\end{align}

As photons interact with the target coherently, one can not determine the 
positions of photoproduced $J/\psi$s exactly. In order to consider 
the modifications of anisotropic QGP on photoproduced mesons, we need the spatial 
distributions of photoproduced $J/\psi$s. 
$J/\psi$s are mainly produced by Pomeron exchange process, we distribute 
the photoproduced $J/\psi$s over the nucleus surface with a 
normalized distribution $f^{norm}({\bf x}_T)$ proportional to the 
square of target thickness function $\sim T_{A}^2({\bf x}_T)$. 
The other limit 
where photoproduced $J/\psi$s are \emph{uniformly} 
distributed over the target nuclear surface 
will be discussed below, to check the effects of this 
distribution uncertainties 
on the final results. 

Now we supplement QGP modifications on photoproduced charmonium, 
{ \small{
\begin{align}
{d\widetilde{N}_{J/\psi}
\over dy}(y|b) = \int d^2{\bf x}_T 
%N_{J/\psi}^{(1)} 
&w n_\gamma(w,b)\sigma_{\gamma A\rightarrow J/\psi A}(w) 
\times
f^{norm}({\bf x}_T+{\bf b}/2)\times [\mathcal{R}_g({\bf x}_T+{\bf b}/2,x,\mu)]^2
\times e^{-\int_{\tau_0}^{\tau_f} d\tau 
\alpha_{\Psi}({\bf x}_T,{\bf b}, \tau) } \nonumber \\
&+(y\rightarrow -y, {\bf b}/2\rightarrow -{\bf b}/2)
\end{align}} }
where $\tau_f$ is the lifetime of QGP. 
We set the origin of coordinates at the middle of the centers of nuclei, 
and two nuclear centers locate at the positions 
of ($0, \pm b/2$) respectively in transverse plane.  
$J/\psi$ decay rate in hot medium $\alpha_{\Psi}$ is given 
in Eq.(\ref{Cdfactor}). Equivalent photon density and photon-nucleus cross section 
are derived below.

Based on EPA method, we obtain the the photon spectrum by setting   
the energy flux of the fields equal to the 
energy flux of equivalent photons through the transverse plane,   
\begin{align}
\int_{-\infty}^{\infty} d\tau\int d{\bf x}_T\cdot {\bf S}=\int_0^{\infty} dww 
{n(w)}
\label{eq-photonden}
\end{align}
where ${\bf S}={\bf E}_T\times {\bf B}_T$ is 
the Poynting vector and $n_\gamma(w)\equiv dN_\gamma/dw$ is 
the photon spectrum. Introducing the impact parameter dependence in photon 
spectrum, it becomes~\cite{Vidovic:1992ik},
\begin{align}
n_\gamma(w,b)&= {1\over \pi w} |{\bf E}_T(w,b)|^2 
\nonumber \\
&={Z^2\alpha\over 4\pi^3 w}|\int_0^\infty dk_T k_T^2 
{F(k_T^2 +{w^2\over \gamma_L^2})
\over k_T^2 +{w^2\over \gamma_L^2}}J_1(bk_T)|^2
\label{eq-spectrum}
\end{align}
where $Ze$ is the nuclear charge ($e$ is the magnitude of electron charge). 
Here we employ the relation of 
$|{\bf E}_T|\approx |{\bf B}_T|$ in the first line 
in the limit of $v/c\approx 1$, $v$ is the nucleon velocity. 
Nucleon Lorentz factor $\gamma_L=\sqrt{s_{NN}}/(2m_N)$ is at the orders of 
$\sim 100$ at RHIC and $\sim 1000$ at LHC.  
$\alpha=e^2/(\hbar c)$ is the electromagnetic coupling constant ($\hbar =c=1$ in 
this work). 
$k_T$ is the photon transverse momentum, 
$J_1$ is the first kind Bessel function.
The nuclear form factor $F(q^2)$ is the Fourier transform 
of the charge distrbution in nucleus (taken as Woods-Saxon distribution).
In the Ultra-peripheral collisions with $b>2R_A$,  
one can approximate the nucleus as a pointlike particle without inner charge 
distribution, and still obtain similar 
electromagnetic fields~\cite{Klusek-Gawenda:2015hja}. With the form factor of 
a pointlike particle, the photon spectrum Eq.(\ref{eq-spectrum}) is given 
analytically, 
\begin{align}
n_\gamma(w,b)={Z^2w\alpha\over 4\pi^3\gamma_L^2}[K_1({wb\over \gamma_L})]^2
\label{eq-point}
\end{align}
where $K_1$ is a modified Bessel function. 
Note that this photon spectrum Eq.(\ref{eq-point}) for
pointlike particle is close to the realistic situation at UPCs, 
but diverges at small 
impact parameter $b<2R_A$ where nucleus can not be treated as 
a ``pointlike'' particle anymore. 
In this work, we do the realistic calculations 
with Eq.(\ref{eq-spectrum}) in all centralities. For the comparision between 
different form factors, please see Ref.\cite{Klusek-Gawenda:2015hja}. 

There is another ingredient we need for the charmonium photoproduction, 
$\sigma_{\gamma A\rightarrow J/\psi A}(w)$, which can be obtained from 
photon-proton cross sections with optical theorem and Glauber calculations. 
We write photo-nuclear cross section into a differential form, 
\begin{align}
\sigma_{\gamma A\rightarrow J/\psi A}= {d\sigma_{\gamma A\rightarrow J/\psi A}
\over dt}|_{t=0} \int_{-t_{min}}^\infty dt |F(t)|^2
\end{align}
Here $-t_{min}=[M_{J/\psi}^2/(4w\gamma_L)]^2$ is 
the minimum momentum transfer 
squared needed to produce a vector meson of mass $M_{J/\psi}$ in the 
laboratory reference frame. 
Note the nuclear form factor $F(t)$ is significant only for $|t|<(1/R_A)^2$, 
The differential photo-nuclear cross section is 
\begin{align}
\label{eq-JpsiA}
&{d\sigma_{\gamma A\rightarrow J/\psi A}\over dt}|_{t=0} 
= {\alpha \sigma_{tot}^2(J/\psi A)\over 4f_V^2} \\
\label{eq-Jpsip}
&\sigma_{tot}(J/\psi A) = \int d^2{\bf x}_T [1-e^{-\sigma_{tot}(J/\psi p)T_A(
{\bf x}_T)}]
\end{align}
where $f_V^2/4\pi =10.4$ is determined by $J/\psi$ mass 
and its leptonic decay 
width with a proper modification based on 
generalized vector dominance model (GVDM) and 
photoproduction data~\cite{Pautz:1997eh}. 
Eq.(\ref{eq-Jpsip}) express the $J/\psi$-nuclear cross section as a function 
of $J/\psi$-proton cross section by thickness function. 
Using optical theorem again, and connect the vector meson-proton cross section 
with the photon-proton cross section measured by HERA, one can obtain
\begin{align}
\sigma_{tot}^2(J/\psi p) = {4f_V^2\over \alpha} {d\sigma(\gamma p\rightarrow 
J/\psi p)\over dt}|_{t=0}
\end{align}
There are Pomeron and meson exchange portion contributing to the cross section. 
The meson exchange term falls rapidly as the center of mass energy 
$W_{\gamma p}$ increases, it is strongly suppressed in heavy vector meson production 
such as $\phi$ and $J/\psi$. The cross section may be parametrized with HERA data 
as~\cite{book:HERA}  
\begin{align}
{d\sigma(\gamma p\rightarrow J/\psi p) \over dt}|_{t=0} = b_V X W^\epsilon 
\times [1-({M_{J/\psi}+M_p\over W_{\gamma p}})^2] 
\end{align}
with $b_V=4\ \rm{GeV^{-2}}$, $X=0.00406\ \mu b$ and $\epsilon=0.65$. 
$M_{J/\psi}$ and $M_p$ are the mass of $J/\psi$ and proton. 
The center of mass energy of the scattering particles ($\gamma$ and $p$) is 
$W_{\gamma p}=2\sqrt{wE_p}$ with $E_p$ to be the proton energy. 
For more details about the extraction 
of $\gamma p$ cross section, please see~\cite{Klein:1999qj}. 

The vector mesons are produced after photon fluctuating into hadronic states 
through two gluon exchanges with the target nucleus. Therefore, vector 
meson production is proportional to the square of gluon densities in the 
target nucleus. With shadowing effect modifications 
on nuclear gluon distributions at small Bjorken-$x$, 
photoproduction of vector mesons 
can be significantly suppressed especially at LHC energies. 
For photoproduced vector mesons, the resolution scale in gluon density 
is chosen 
as $\mu^2=M_{J/\psi}^2/4\ \rm{GeV^2}$~\cite{Ryskin:1992ui,Guzey:2016piu}. 
With the relation of $-t\ll M_{J/\psi}^2$, 
the fraction of gluon momentum is around $x\approx M_{J/\psi}^2/W_{\gamma p}^2
={M_{J/\psi}\over \sqrt{s_{NN}}}e^{-y}$, at the order of $10^{-3}\sim 10^{-5}$ at 
$\sqrt{s_{NN}}=2.76$ TeV. 

With charmonium hadroproduction (initial production, regeneration and 
decay from B hadrons) and photoproduction, now we can write the prompt 
and inclusive nuclear 
modification factors, 
\begin{align}
\label{eqRAAprom}
&R_{AA}^{\rm prompt} = {N^{\gamma A\rightarrow J/\psi A} +N_{AA}^{J/\psi} \over 
N_{pp}^{J/\psi}N_{coll}} \\
\label{eqRAAincl}
&R_{AA} ={R_{AA}^{\rm prompt}\over 1+r_B} +{r_B\times Q\over 1+r_B}
\end{align}
where $N^{\gamma A\rightarrow J/\psi A}$ represents 
the charmonium photoproduction with 
hot medium effects in AA collisions. Photoproduction in pp 
collisions is negligible compared with AA situation. 
$N_{AA}^{J/\psi}$ is $J/\psi$ prompt hadroproduction, obtained from the 
integration of Eq.(\ref{eq-analy}). 
$N_{coll}=\sigma_{incl}
\int d^2{\bf x}_T T_A({\bf x}_T+{\bf b}/2) T_B({\bf x}_T-{\bf b}/2)$ 
is the number 
of binary collisions, with $\sigma_{incl}$ to be the inelastic cross section 
between two nucleons. $r_B=f_B/(1-f_B)$ is the ratio of non-prompt 
over prompt $J/\psi$ yields in pp collisions. 
In a certain $p_T$ window, 
bottom quark number can be modified 
by a quench factor $Q= N_{AA}^{b}/(N_{coll}N_{pp}^{b})$ in QGP 
due to the energy loss, which satisfies $Q\ge 1$ and $Q\le 1$ in small and 
high $p_T$ regions, respectively~\cite{Chen:2013wmr}. 
Considering that the 
fraction of non-prompt $J/\psi$ is much smaller than the prompt part, 
it will change $J/\psi$ inclusive $R_{AA}$ slightly and is neglected in this 
work by taking $Q=1$. 
Note that in the situation where B hadron decay becomes dominant such as 
$\psi(2S)$ inclusive yields in Pb-Pb collisions, 
effects of bottom quark energy loss should be treated 
seriously~\cite{Chen:2013wmr}.    

In Ultra-peripheral collisions with $b>2R_A$, there will be no hadroproduction and 
denominator of Eq.(\ref{eqRAAprom}) approaches to zero. In the 
extremely low $p_T$ regions where photoproduction becomes important, 
photoproduction becomes non-zero at the limit of 
$N_p\rightarrow 0$, 
makes $R_{AA}^{\rm prompt}\rightarrow 
\infty$ and $R_{AA}\rightarrow \infty$, see both experimental data and 
our theoretical results below. 

\begin{figure}[htb]
{\includegraphics[width=0.47\textwidth]{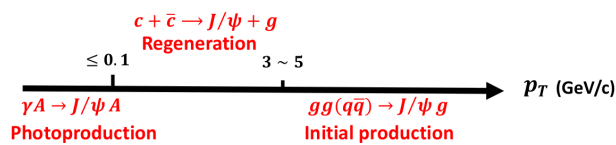}
\caption{
 (Color online) Schematic diagram for 
different charmonium production mechanisms 
at different transverse momentum regions 
in semi-central nucleus-nucleus collisions 
with the existence of both QGP and strong transverse electromagnetic field. 
Photoproduction, regeneration and initial production dominates $J/\psi$ final 
yields in extremely low $p_T$, low and middle $p_T$, high $p_T$ regions, 
respectively.
}
\label{fig-mech}}
\end{figure}

Before quantitative calculations, let's analyse the 
different charmonium production in different $p_T$ regions. 
As equivalent photon number decreases with frequency rapidly, 
photoproduced $J/\psi$s mainly distribute at $p_T<0.1$ GeV/c with mean 
transverse momentum $\langle p_T\rangle_{J/\psi}\approx 
0.055$ GeV/c~\cite{Adam:2015gba}. Beyond this $p_T$ bin, the 
hadroproduction becomes important. 
With strong energy loss of charm quarks, 
the regenerated $J/\psi$s are 
mainly distributed below $3\sim 5$ GeV/c. 
At higher $p_T$ bins, $J/\psi$s from 
initial production and decays of B hadrons dominates. 
Please see Fig.\ref{fig-mech}. 

\begin{figure}[htb]
{\includegraphics[width=0.45\textwidth]{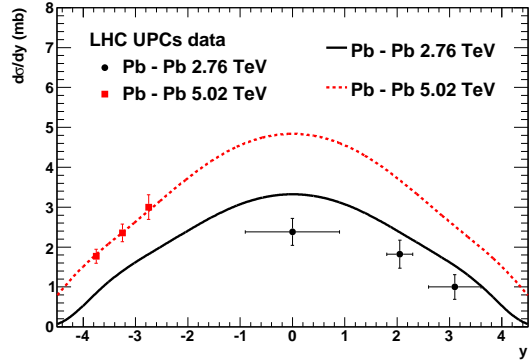}
\caption{(Color online) 
$J/\psi$
coherent photoproduction as a function of rapidity in 
Ultra-peripheral collisions at 2.76 TeV and 5.02 TeV Pb-Pb collisions. 
Theoretical calculations does not include the shadowing effect. 
Experimental data is from ALICE Collaboration~\cite{Abelev:2012ba,Abbas:2013oua,ALICE:pre}. 
}
\label{fig-UPC}}
\end{figure}

Before starting the comparison between charmonium hadroproduction and 
photoproduction in semi-central collisions with the existence of QGP, 
we give the charmonium photoproduced cross section as a function of rapidity, 
which has been well studied in 
Ultra-peripheral collisions. With the increasing rapidity, it takes more 
energy $w$ for quasi-real photons 
to fluctuate into a $J/\psi$ with rapidity $y$, $w\propto e^y$. Therefore, 
the cross section in Fig.\ref{fig-UPC} decreases with rapidity. 
After considering the shadowing effect, both lines in Fig.\ref{fig-UPC} will be 
shifted downward, with the magnitude depending on different 
shadowing suppression factors. 

\begin{figure}[htb]
{\includegraphics[width=0.45\textwidth]{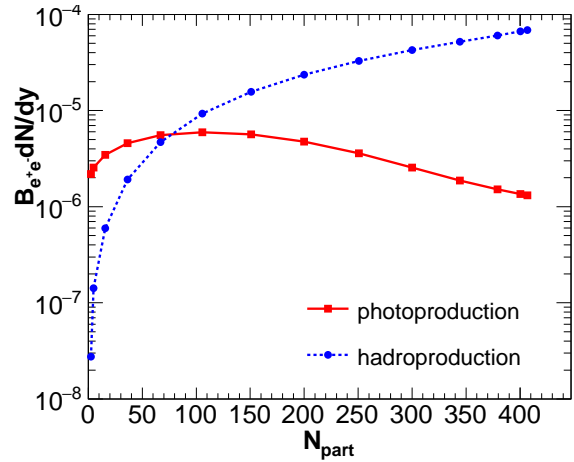}
\caption{ (Color online) 
Charmonium hadroproduction scaled from $pp$ collisions 
and photoproduction as a function of number of participants $N_p$ at the 
forward rapidity $2.5<y<4$ in 
$\sqrt{s_{NN}}=2.76$ Pb-Pb collisions in the extremely low transverse 
momentum region $p_T<0.3$ GeV/c. Both lines are without modifications 
from shadowing effect and QGP suppression. $B_{e^+e^-}$ is the branch ratio 
of $J/\psi\rightarrow e^+e^-$. 
}
\label{fig-initphoto}}
\end{figure}

Charmonium hadroproduction is proportional 
to the number of binary collisions $N_{coll}$. 
With increasing overlap area between two nuclei, the number of binary 
collisions and participants increase significantly. 
However, the electromagnetic fields produced by two nuclei increases at first 
and then decreases with $N_p$, which makes photoproduction reach the 
maximum value at semi-central collisions. In Fig.\ref{fig-initphoto}, 
the solid and dotted lines represent photoproduction and initial 
hadroproduction scaled from \emph{pp} collisions, both 
without modifications from cold (such as shadowing effect) and hot medium 
effects. It helps to clarify the relation between two production mechanisms.   
At around $N_p>200$ (with $b< 7$ fm), the hadroproduction becomes about 
five times larger than the photoproduction, and the interactions between 
strong electromagnetic fields and target nucleus seems negligible. 
With the fact that charmonium 
from hadronic collisions only distributed in the overlap area of two 
nuclei where QGP is produced, but coherently photoproduced charmonium is
distributed over the entire target nucleus surface, 
the QGP suppression on photoproduced charmonium should always be weaker than 
the hadroproduction. 
At $N_p$ of $100\sim150$ where QGP initial temperature is $\sim 2T_c$, 
$20\%\sim 40\%$ of photoproduced $J/\psi$s is dissociated due to parton 
inelastic collisions and color screening effect. 

\begin{figure}[htb]
{\includegraphics[width=0.45\textwidth]{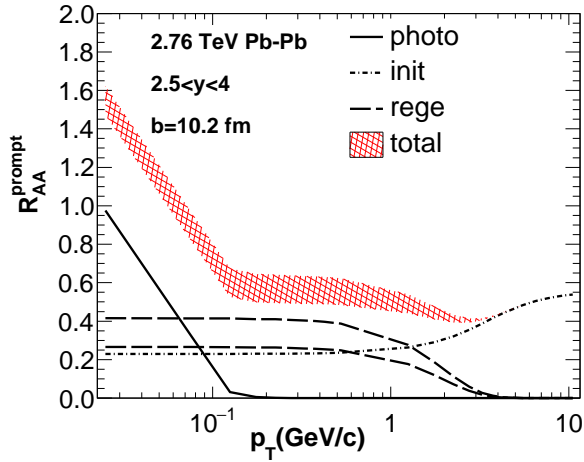}
\caption{(Color online) 
Charmonium \emph{prompt} nuclear modification 
factor as a function of transverse momentum, at 
the impact parameter $b=10.2$ fm in forward rapidity $2.5<y<4$ at LHC 2.76 
TeV Pb-Pb collisions. Dot-dashed line is initial production, dashed lines 
is the regeneration (two lines represent $20\%$ difference of charm pair 
production cross section $d\sigma_{pp}^{c\bar c}/dy$ due to its large 
uncertainties measured by experiments). Solid line is the photoproduction with 
cold and hot medium modifications. Color band is the total contribution (including 
initial production, regeneration and photoproduction).  
}
\label{fig-RAApT}}
\end{figure}

Coherent photoproduced charmonium is from the quasi-real photon fluctuations into 
a vector meson, which is different from the parton fusions in hadronic collisions. 
In Fig.\ref{fig-RAApT}, we compare the different mechanisms for charmonium 
production as a function of $p_T$ with impact parameter b=10.2 fm. 
The hadroproduction shows a flat behavior in $p_T<1$ GeV/c. In the $p_T<0.1$ GeV/c 
where photoproduction becomes non-zero (solid line), 
$R_{AA}$ is significantly enhanced and 
even larger than the unit. With increasing $p_T$, 
charm quark density in QGP decreases as a fermi-distribution 
(thermalization limit) or power law (given by pQCD), regeneration (dashed line) 
drops to zero 
at $p_T\sim 3$ GeV/c~\cite{Zhao:2017yan}. In higher $p_T$ bins, 
final charmonium mainly consists of initial production from parton 
hard scatterings. The tendency of 
$R_{AA}(p_T)$ in Fig.\ref{fig-RAApT} is very similar with RHIC data at 
cent.20-40\%~\cite{Zha:2017etq}. The enhancement of 
$R_{AA}$ in $p_T<0.1$ GeV/c will be more significative at larger  
impact parameter. 
 
\begin{figure}[htb]
{\includegraphics[width=0.45\textwidth]{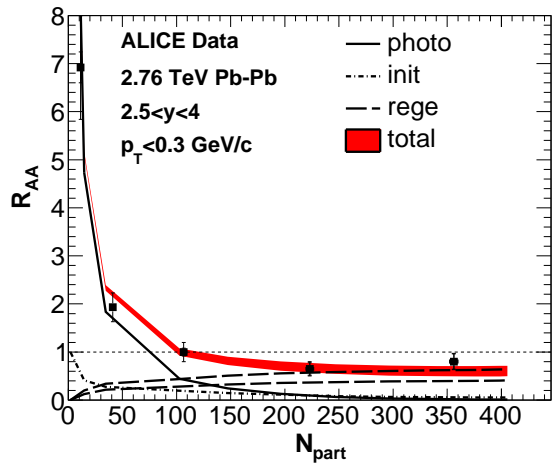}
\caption{(Color online) 
Charmonium \emph{inclusive} nuclear modification factor as a function of number of 
participants $N_p$ in $p_T<0.3$ GeV/c in forward rapidity $2.5<y<4$ at LHC 2.76 
TeV Pb-Pb collisions. (Dot-dashed, dashed, solid) lines are the 
initial production, regeneration and photoproduction, respectively. 
Color band is the total production including decay from B hadrons 
(non-prompt $J/\psi$). Experimental data is from 
ALICE Collaboration~\cite{Adam:2015gba}. 
}
\label{fig-RAANp}}
\end{figure}

\begin{figure}[htb]
{\includegraphics[width=0.43\textwidth]{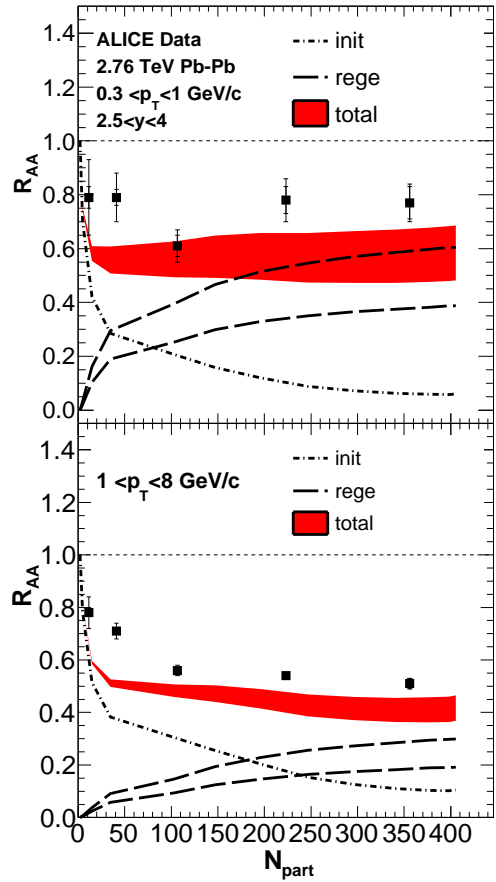}
\caption{(Color online) 
Charmonium \emph{inclusive} nuclear modification factors as a function of number of 
participants $N_p$ in $0.3<p_T<1$ and $1<p_T<8$ GeV/c 
in forward rapidity $2.5<y<4$ at LHC 2.76 
TeV Pb-Pb collisions. 
Lines and color band are the same with Fig.\ref{fig-RAANp}. The upper limit of 
color band will be shifted upward if employing a larger charm cross section 
$d\sigma_{c\bar c}^{pp}/dy$ which is with large uncertainty in 
experiments~\cite{Abelev:2012vra}. Note that 
photoproduction in these $p_T$ regions is negligible and therefore absent in 
this figure. Experimental data is from ALICE Collaboration~\cite{Adam:2015gba}. 
}
\label{fig-RAANp2}}
\end{figure}

Now we give the nuclear modification factor as a function of centrality. 
The experimental data in Fig.\ref{fig-RAANp}-\ref{fig-RAANp2} is for 
inclusive $J/\psi$, we already 
include the contribution from B hadron decay in the total production (color band).
In the central collisions, the regeneration becomes important due to the abundant 
number of $c\bar c$ pairs in QGP. In peripheral collisions, both QGP lifetime 
and charm pair number become smaller, which suppresses the regeneration. 
The initially produced $J/\psi$s suffer weaker suppression in the semi-central and 
peripheral collisions. Photoproduction from strong electromagnetic fields 
dominates in the peripheral collisions. 
The nuclear modification factor $R_{AA}$ in $p_T<0.3$ GeV/c (Fig.\ref{fig-RAANp}) 
is much larger than the value in $p_T>0.3$ GeV/c (Fig.\ref{fig-RAANp2}) at 
$N_p\le 100$. This $R_{AA}$ enhancement 
in particular $p_T$ bin 
is attributed to the photoproduction. 


Finally, let's analyse the uncertainties in our theoretical calculations. 
The photon-nucleus cross section $\sigma_{\gamma A\rightarrow J/\psi A}$ depends 
on the parametrization of $\gamma p$ differential cross section and 
charmonium properties~\cite{Santos:2014zna}. 
This uncertainty can be partially constrained 
by the experimental data of charmonium photoproduction 
in UPCs. 
One of another important inputs is the 
charmonium hadronic cross section $d^2\sigma_{pp}^{J/\psi}/dydp_T$ (
used in the numerator and denominator of $R_{AA}$) in the 
low $p_T$ region absent of experimental data. Our parametrization 
Eq.(\ref{eq-ppJpsi}) is very close to the 
parametrization employed by ALICE Collaboration to 
obtain experimental $R_{AA}$, with the difference of $p_T$-dependence 
less than $4\%$~\cite{Bossu:2011qe}, which 
ensures the same definitions of 
our theoretically calculated $R_{AA}$ and the experimental data.  
Another uncertainty is 
the shadowing effect. With EPS09 model, 
it suppresses gluon distribution by 
around 20\% in the Bjorken-$x$ corresponding 
to charmonium in forward rapidities. However, 
different models give very different magnitudes 
of shadowing effect. In the limit without shadowing effect, both charmonium 
photoproduction 
and hadroproduction become $\sim1/0.8^2$ times of above results. 
For the QGP induced $J/\psi$ decay rate, our values from Eq.(\ref{Cdfactor}) 
is similar with other groups~\cite{Zhao:2010nk}.  
In order to introduce the spatial distribution of coherently 
photoproduced $J/\psi$ over the 
nuclear surface, we assume the charmonium density to be proportional to the 
square of gluon densities inspired by the Pomeron exchange process. 
In the other limit that 
photoproduced charmonium distribution is proportional to the photon 
spatial density which 
changes slowly with the coordinates, we assume an 
\emph{uniform} 
charmonium distribution over the nuclear surface. 
With the fact that shadowing effect becomes weaker in the edge of nucleus, 
\emph{uniform} distribution suffers weaker shadowing effect and QGP dissociation. 
These make the final photoproduction $N^{\gamma A\rightarrow J/\psi A}$ in 
Eq.(\ref{eqRAAprom}) increases by around 20\% in semi-central collisions. 

The incoherently photoproduced $J/\psi$s 
can also contribute to the charmonium $R_{AA}$. They are mainly distributed 
in a larger $p_T$ region compared with coherent $J/\psi$, 
such as $0.2<p_T<0.8$ GeV/c~\cite{Abelev:2012ba}. 
In $p_T<0.2$ GeV/c, incoherent production 
is much smaller than the coherent part, and neglected in our 
calculations. In higher $p_T$ regions, it becomes the main source for $J/\psi$ 
producion in Ultra-peripheral collisions. 
However, in semi-central collisions where hadroproduction 
increases linearly with $p_T$ at $p_T<1$ GeV/c~\cite{Zhao:2017yan}, 
incoherent photoproduction is also negligible for the $p_T$-integrated observables 
at $0.3<p_T<1$ GeV/c. For the particular $p_T$ bin, 
such as $0.1\sim 0.5$ GeV/c, the importance of incoherent 
photoproduction needs detailed studies.

In summary, we employ the Boltzmann-type transport model to study 
charmonium hadronic production including initial production, regeneration 
and decay from B hadrons. In extremely low $p_T$ region, charmonium 
photoproduction is supplemented with 
modifications of both shadowing effect and QGP suppressions. 
In peripheral collisions and extremely small $p_T$ region, 
photoproduction 
becomes larger than the hadroproduction, and is mainly suppressed by 
shadowing effect with large uncertainties instead of QGP. 
However, in semi-central 
collisions, both shadowing effect and QGP suppression becomes important 
on charmonium production. And around $20\%\sim 40\%$ of photoproduced 
$J/\psi$s can be dissociated by QGP at $100<N_p<150$. In 
(extremely low, middle, high) $p_T$ regions, 
photoproduction, regeneration 
and initial production play the dominant roles respectively 
in charmonium final production,  
which help to clarify \emph{different physics} such as shadowing effect 
on gluon distribution 
in small Bjorken-$x$, charm quark thermalization (controls regeneration), 
charmonium dissociation by QGP in relativistic heavy ion collisions.

\vspace{1cm}
\appendix {\bf  Acknowledgement}: 
We thank Guansong Li for helpful discussions. 
WS and BC are supported by NSFC Grant No. 11705125 and 11547043, 
BC is also supported by Sino-Germany (CSC-DAAD) Postdoc Scholarship. 
WZ is supported by NSFC Grant No. 11775213 and 11505180.

\end{document}